\documentclass[12pt]{article}
\usepackage{epsfig}
\usepackage{psfrag}
\usepackage{enumitem}
\usepackage{latexsym}
\usepackage{indentfirst}
\usepackage{fancyhdr}
\usepackage{amssymb}
\usepackage{amsmath}
\usepackage{amsfonts}
\usepackage{pifont}
\usepackage{cite}
\usepackage{bbold}
\usepackage{color}
\usepackage{graphics}
\usepackage[center,footnotesize,hang]{subfigure}
\usepackage{url}
\usepackage{array}

\def\be{\begin{equation}}
\def\ee{\end{equation}}
\def\bc{\begin{center}}
\def\ec{\end{center}}
\def\bea{\begin{eqnarray}}
\def\eea{\end{eqnarray}}

\newcommand{\ba}{\begin{array}{c}}
\newcommand{\bad}{\begin{array}{ccc}}
\newcommand{\ea}{\end{array}}

\def\nn{\nonumber}

\begin{document}

\begin{titlepage}
\hfill{RM3-TH/12-11}
  \vskip 2.5cm
  \begin{center}
   {\Large\bf   Leptonic CP violation at neutrino telescopes}
   \end{center}
  \vskip 0.2  cm
  \vskip 0.5  cm
   \begin{center}
   {\large Davide Meloni}~\footnote{e-mail address: davide.meloni@fis.uniroma3.it}
   \\
   \vskip .2cm {\it Dipartimento di Fisica ``E.~Amaldi'',}
   \\
   {\it Universit\'a degli Studi Roma Tre,\\ Via della Vasca Navale 84, 00146 Rome, Italy}
   \\
   \end{center}
   \begin{center}
   {\large Tommy Ohlsson}~\footnote{e-mail address:  tohlsson@kth.se}
   \\
   \vskip .2cm {\it Department of Theoretical Physics, School of
Engineering Sciences,\\
KTH Royal Institute of Technology -- AlbaNova University Center,}
   \\
   {\it Roslagstullsbacken 21, 106~91~Stockholm, Sweden}
   \\
   \end{center}
  \vskip 0.7cm
 
 \begin{abstract}
 \noindent
With the advent of the recent measurements in neutrino physics, we investigate the role of high-energy neutrino flux ratios at neutrino telescopes for the possibility of determining the leptonic CP-violating phase  $\delta$ and the underlying pattern of the leptonic mixing matrix. 
We find that the flux ratios show a dependence of ${\cal O}(10~\%)$ on 
the CP-violating phase,  and for optimistic uncertainties on the flux ratios less than 10~\%, they can be used to distinguish between CP-conserving and 
CP-violating values of the phase at 2$\sigma$ in a non-vanishing interval around the maximal value $|\delta|=\pi/2$. 
 \end{abstract}
 \end{titlepage}
 \setcounter{footnote}{0}
 \vskip2truecm

\section{Introduction}

Ever since 1998 with the results of Super-Kamiokande on atmospheric neutrinos \cite{Fukuda:1998mi}, there is strong evidence 
for neutrino oscillations 
as the main mechanism for flavor transitions. Indeed, the fundamental parameters such as the two neutrino mass-squared 
differences $\Delta m_{31}^2$ and 
$\Delta m_{21}^2$ as well as the three leptonic mixing angles $\theta_{23}$, $\theta_{12}$, and $\theta_{13}$ are now determined with increasing accuracy. Recently, 
the third and last mixing angle $\theta_{13}$ has been measured by Daya Bay \cite{An:2012eh}, but also Double Chooz, MINOS, RENO, and T2K have made important 
contributions \cite{Abe:2011fz,Adamson:2011qu,Ahn:2012nd,Abe:2011sj}. All five experiments indicate that the value of the third mixing angle is around nine degrees. 
Thus, the value of this mixing angle 
is relatively large. This means that both the bi- and tri-bi-maximal mixing patterns for leptonic flavor mixing have essentially been ruled out as 
zeroth-order approximations. Therefore, appropriate corrections must be taken into account to reconcile them 
with the experimental data.
To this end, there remain two quantities for massive and mixed neutrinos that can be determined with oscillations: ${\rm sgn}(\Delta m_{31}^2)$ and the Dirac CP-violating phase $\delta$. 
The latter always appears in combination with $\theta_{13}$ in the leptonic mixing matrix as $\sin(\theta_{13}) \exp(\pm {\rm i} \delta)$ and a non-zero value of $\theta_{13}$ means that it is possible to determine $\delta$. The best options to measure $\delta$ are provided by accelerator experiments (e.g.~NOvA and NuMI), future superbeams, 
beta beams, or even better a neutrino factory \cite{Choubey:2010zz}. However, it will take a long time before such experiments are realized. 
In this work, we will therefore consider an alternative approach to determine $\delta$ by measuring neutrino flux ratios at neutrino telescopes. 
The generic example of such a telescope is IceCube \cite{icecube}, which exists and has the potential to measure flux ratios \cite{Lipari:2007su}.
The dependence of flux ratios on the mixing parameters has been addressed in the literature where the focus has been 
on the dependence on $\theta_{23}$ and $\theta_{13}$ \cite{Meloni:2006gv,th23,mutau}, on $\delta$ \cite{delta}, and on new physics effects \cite{newphys}.

The aim of this work is to illustrate the potential of neutrino telescopes to detect $\delta$ through the measurement of flux ratios after the first measurements of $\theta_{13}$. 
For our numerical evaluations, we adopt the best-fit values and uncertainties for the mixing angles 
in the normal hierarchy (NH) 
quoted in Ref.~\cite{Tortola:2012te}, 
which are presented in Tab.~\ref{mixings}.
Note that, although the flux ratios are not sensitive to the mass-squared differences, a somewhat dissimilar behavior between NH and 
inverted hierarchy (IH) could be drawn if 
the allowed ranges for the parameters were sizeably different. However, this is not the case for the  3$\sigma$ ranges considered in this analysis.
\begin{table}[h!]
\centering
\begin{tabular}{l|c|c|c|c}
\it Parameter & \it Best-fit value & $1\sigma$ \it range & $2\sigma$ \it range & $3\sigma$ \it range \\
\hline
$\sin^2 \theta_{12}/10^{-1}$   & 3.2 & 3.03 -- 3.35 & 2.9 -- 3.5 & 2.7 -- 3.7 \\
\hline 
$\sin^2 \theta_{13}/10^{-2}$  & 2.6 & 2.2 -- 2.9 & 1.9 -- 3.3 & 1.5 -- 3.6 \\
\hline 
$\sin^2 \theta_{23}/10^{-1}$  & 4.9 & 4.4 -- 5.7 & 4.1 -- 6.2 & 3.9 -- 6.4 \\
\hline 
$\delta/\pi$  & 0.83 & 0.19 -- 1.37 & $[0,2\pi]$
&$[0,2\pi]$ \\
\end{tabular}
\caption{\label{mixings}\it  Results of a global analysis \cite{Tortola:2012te} in terms of best-fit values,  
$1\sigma$, $2\sigma$, and $3\sigma$ ranges for NH only. See also Ref.~\cite{Fogli:2012ua} for another global analysis.}
\end{table}


\section{Neutrino flux ratios}
\label{sec:nfr}

The averaged neutrino oscillations probabilities can be written as (see e.g.~Ref.~\cite{Meloni:2006gv})
\begin{equation}
\langle P_{\alpha\beta} \rangle = \sum_{i=1}^3 |U_{\alpha i}|^2 |U_{\beta i}|^2\,,
\end{equation}
where the quantities $U_{\alpha i}$ are elements of the leptonic mixing
matrix $U$. Starting from the flux ratios at a source given by
$\phi_{\nu_e}^0: \phi_{\nu_\mu}^0 : \phi_{\nu_\tau}^0$, the neutrino fluxes
arriving at a detector are sensitive to oscillations in vacuum and
are then computed as
\begin{equation}
\phi_{\nu_\beta} = \sum_{\alpha = e,\mu,\tau} \phi_{\nu_\alpha}^0 \langle
P_{\alpha\beta} \rangle =  \sum_{\alpha = e,\mu,\tau} \phi_{\nu_\alpha}^0 \sum_{i=1}^3 |U_{\alpha i}|^2 |U_{\beta i}|^2\,.
\end{equation}
For example, in the case $\phi_{\nu_e} ^0: \phi_{\nu_\mu}^0 : \phi_{\nu_\tau}^0 = 1:2:0$,
the fluxes are stemming from pion-beam sources ($\pi S$), whereas in the case $\phi_{\nu_e} ^0: \phi_{\nu_\mu}^0 : \phi_{\nu_\tau}^0 = 0:1:0$, 
they come from muon-damped sources ($\mu DS$).
Then, the flux ratios are defined as follows:
\begin{equation}
R_{\alpha\beta} = \frac{\phi_{\nu_\alpha}}{\phi_{\nu_\beta}}.
\end{equation}
Note that the three flux ratios $R_{e\mu}$, $R_{e\tau}$, and
$R_{\mu\tau}$ are not independent of each other, since two of them will give
the third one. For example, using $R_{e\mu} (R_{e\tau})^{-1} R_{\mu\tau} \equiv 1$, we showed in Ref.~\cite{Meloni:2006gv} that, up to second order in small quantities (to be defined below), the following sum-rule holds: 
$R_{e\mu}-R_{e\tau}+R_{\mu\tau} \simeq 1$.
Given also the simplicity of their analytical expressions,  we will choose to examine first the relevant features of $R_{e\mu}$ and $R_{\mu\tau}$.
Notice that a more accessible variable from the experimental point of view is given by
\bea
R=\frac{\phi_{\nu_\mu}}{\phi_{\nu_e}+\phi_{\nu_\tau}} =R_{\mu e} \frac{1}{1+R_{\tau e}} = (R_{e\mu})^{-1} \frac{1}{1 + (R_{e\tau})^{-1}} \,,
\eea
which is obviously related to the ``fundamental'' $R_{\alpha\beta}$. We will take into account $R$ in Section~\ref{exp}, when dealing 
with the issue of achievable precision for the flux ratios at neutrino telescopes. 
It is important to stress that the various flux ratios are experimentally accessible at different neutrino energies.
For example, IceCube has an energy threshold of 100~GeV for detecting
muon tracks, and $\sim$~1~TeV for detecting electron- and tau-related showers. At higher energies, above $\sim$~1~PeV, 
electron-related electromagnetic showers and the tau-related hadronic showers can also be distinguished. 
Here, we do not consider such details and assume to work in the appropriate energy regime, where all relevant $R_{\alpha\beta}$ can be measured.

\section{Investigation of neutrino flux ratios}
\label{sec:invest_nfr}

\subsection{Estimate of the uncertainties for pion-beam sources}

Since the exact formulas for the flux ratios are quite cumbersome, we prefer to present our results 
expanding these ratios in the small parameters $\theta_{13}$, 
$\delta_{23}=\theta_{23}-\pi/4$, and  $\delta_{12}=\theta_{12}-\bar \theta_{12}$, $
\bar \theta_{12}$ being the best-fit value for $\theta_{12}$.
No restrictions have been applied to $\delta$,
which means that the following formulas are valid to all orders in
$\delta$. In our discussion, it is enough to consider the expansions up to 
first order for $R_{e\mu}$ and $R_{e\tau}$ and second order for $R_{\mu\tau}$  
(see Ref.~\cite{Meloni:2006gv} for a detailed discussion),
 we obtain
 \bea
 R_{e\mu} &=& 1+\frac{3}{4} \cos(\delta) \sin(4 \theta_{12}) \, \theta_{13}-\frac{3}{2} \sin^2(2\theta_{12}) \, \delta_{23}\,, \nn\\
 && \label{pion} \\
 R_{\mu\tau} &=& 1
 +2 \cos^2(\delta)\sin^2(2 \theta_{12}) \,
 \theta_{13}^2+2 \cos(\delta) \sin(4 \theta_{12}) \,\theta_{13}\, \delta_{23}+\left[\cos(4 \theta_{12})+7\right] \, \delta_{23}^2\,.
 \nn
 \eea
At this order of perturbation theory, $R_{e\tau}=R_{e\mu}$. Notice also 
that no contributions from $\delta_{12}$ appear at the perturbative level considered above, so we expect that, given the current 
best-fit value on $\theta_{12}$, the impact of $\delta_{12}$ is negligible.
Inserting the best-fit values from Tab.~\ref{mixings} into Eq.~(\ref{pion}), we can evaluate the numerical 
weight of each term in the expansions:
\bea
 \nn
R_{e\mu} &\sim& 1+ 0.5 \cos(\delta) \, \theta_{13}-1.3 \,\delta_{23}  + {\cal O}(\delta_{ij}^2)\,, \nn\\
&& \label{pionnum} \\
R_{\mu\tau} &\sim&  1 + 1.7\cos^2(\delta) \, \theta^2_{13} +  1.3\cos(\delta) \, \theta_{13}\,\delta_{23} +6.3\,\delta_{23}^2. \nn
\eea
Numerically, $\delta_{12}\sim 0.05$,  $\theta_{13}\sim 0.2$, and $\delta_{23}\sim 0.15$.
Since
$\delta_{23}$ always appears with the largest coefficient, we can easily deduce that the main uncertainty comes from the current error on $\theta_{23}$. The contribution of $\theta_{13}$ is modulated by $\delta$, so we expect no impact from it for $\cos(\delta) \sim 0$. 

\begin{figure}[h]
\includegraphics[width=0.45\textwidth]{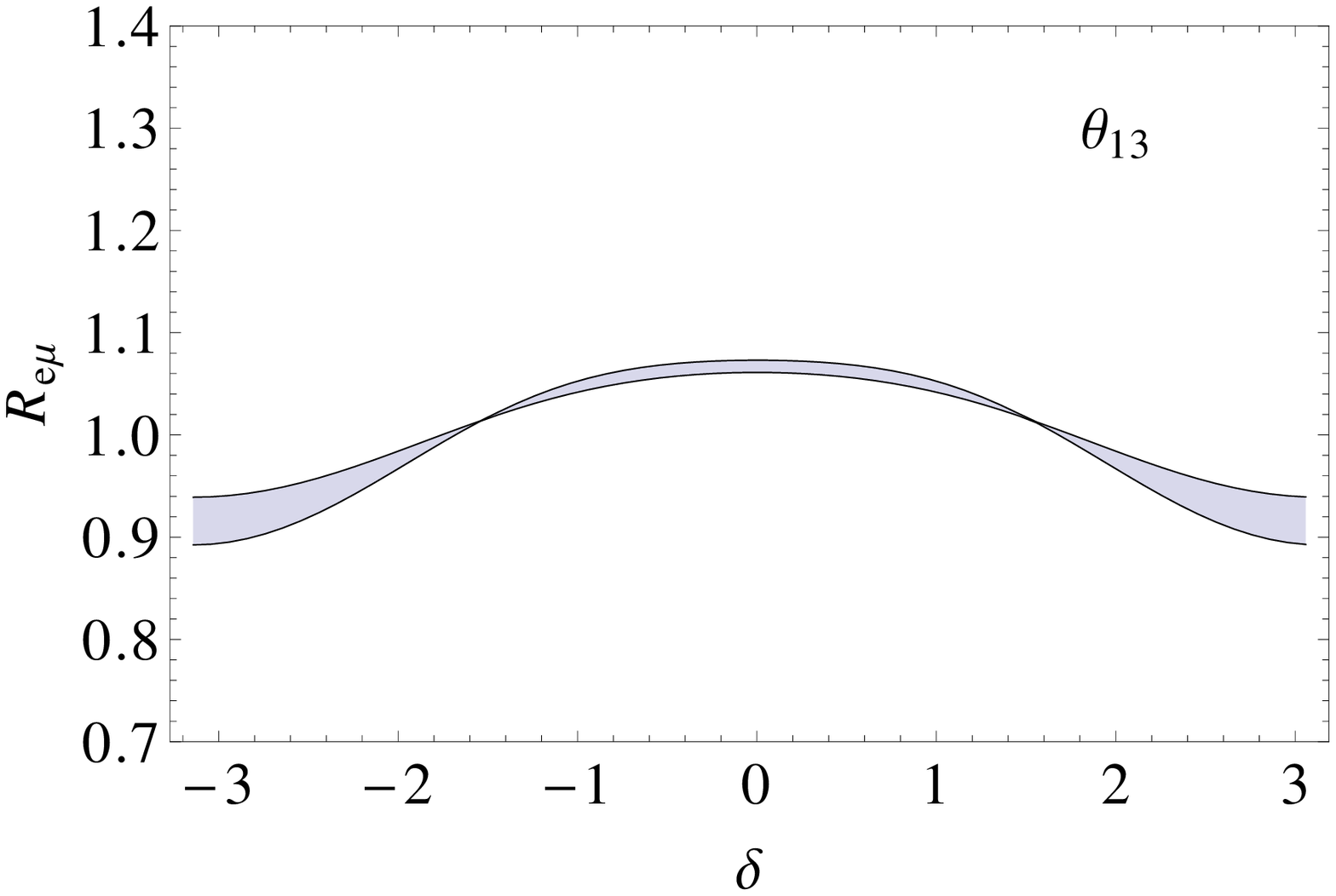}\qquad \includegraphics[width=0.45\textwidth]{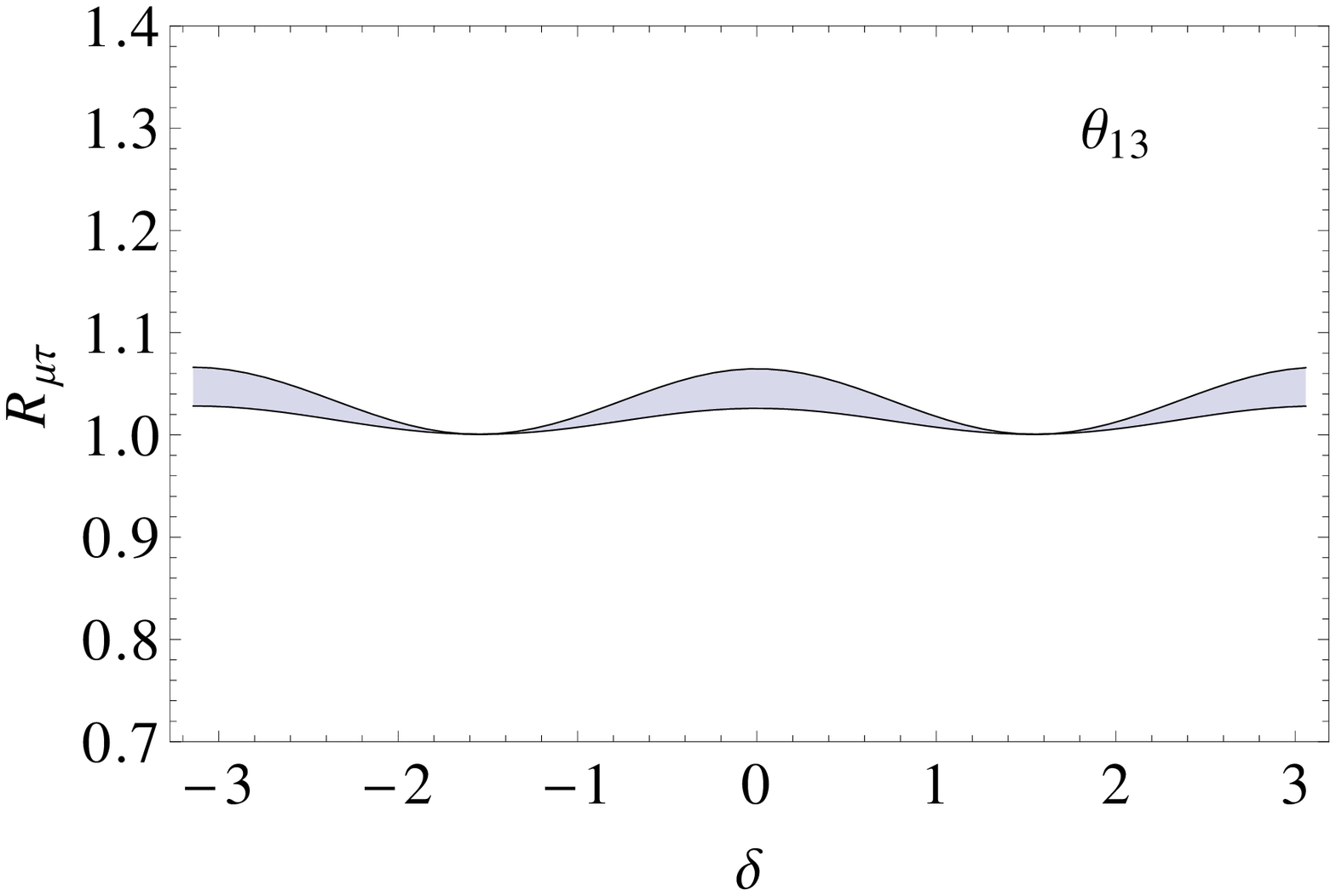} \\
\includegraphics[width=0.45\textwidth]{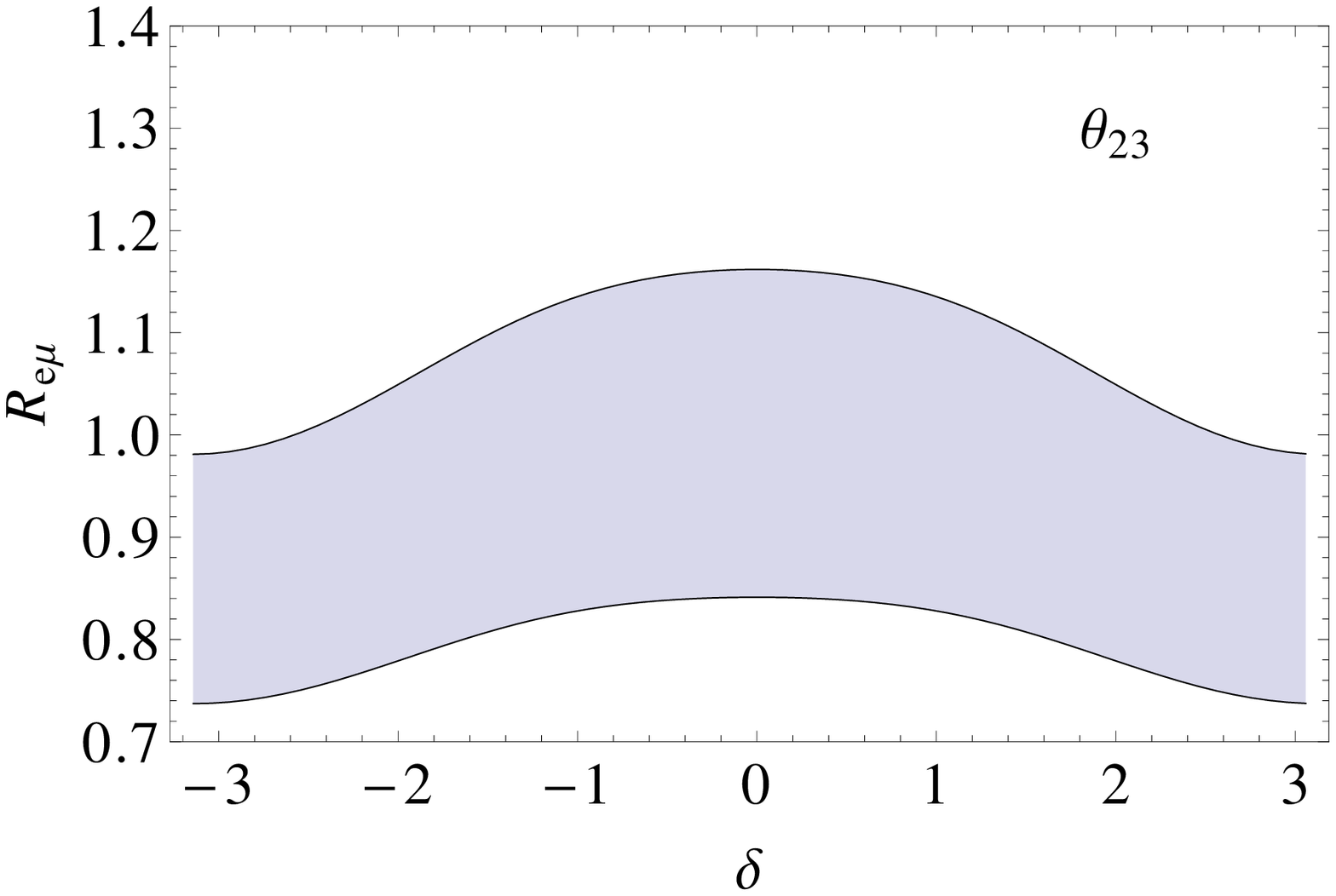}\qquad \includegraphics[width=0.45\textwidth]{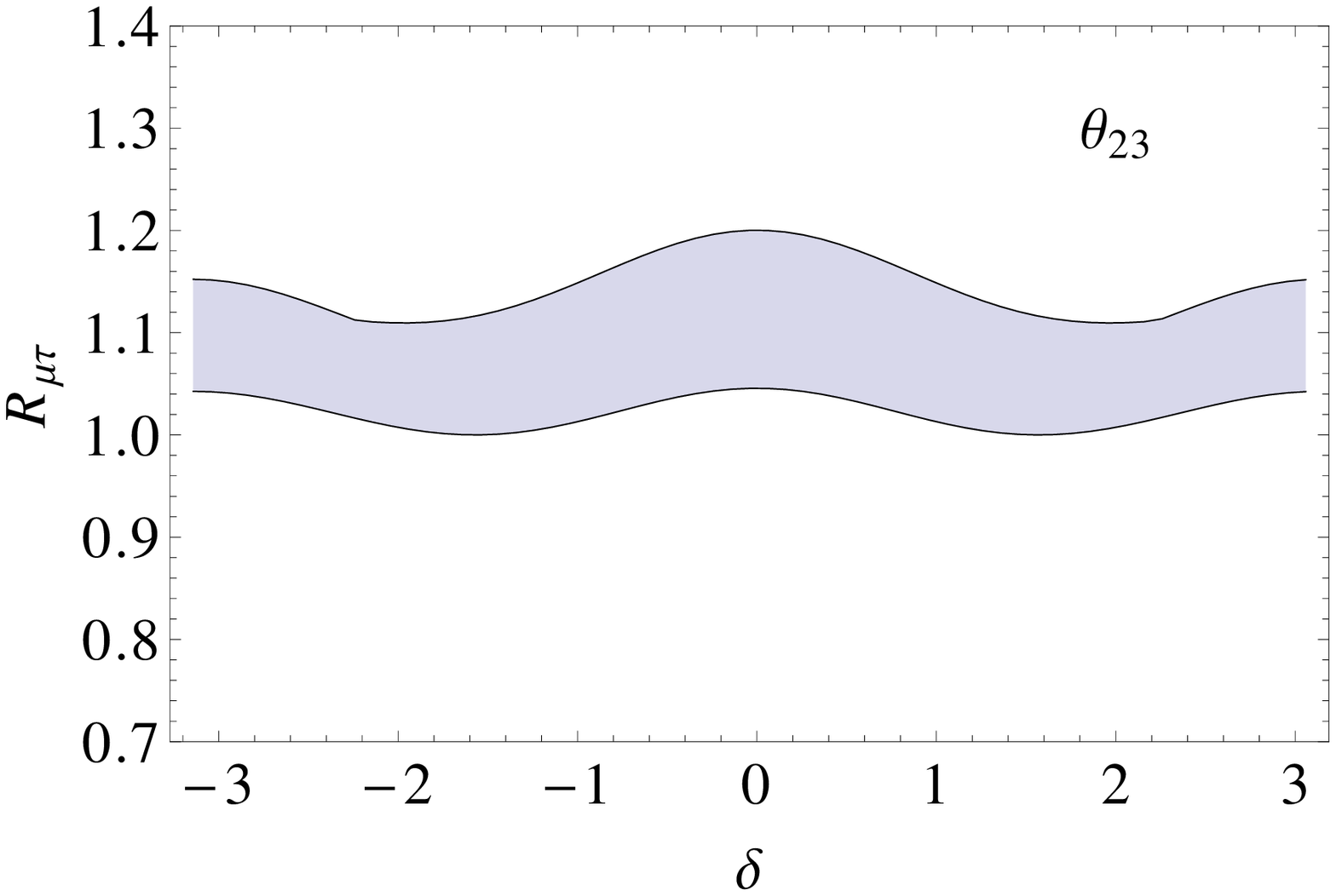} 
\caption{\it Variations of $R_{e\mu}$ (left plots) and $R_{\mu\tau}$ (right plots) as functions of $\delta$ when only 
the uncertainty on $\theta_{13}$ is taken into account (upper plots) or that on $\theta_{23}$ (lower plots).}
\label{fig:unc}
\end{figure}

We verified all these statements by computing the $R_{\alpha\beta}$'s as functions of $\delta$, 
using exact expressions for the flux ratios. 
For the sake of illustration, we present in Fig.~\ref{fig:unc} the results obtained for both $R_{e\mu}$ and $R_{\mu\tau}$.
In the upper plots, we show the behavior of the two flux ratios when $\theta_{13}$ is varied 
inside its 3$\sigma$ range, whereas in the lower plots, we show the same for $\theta_{23}$, with the unshown parameters fixed to their best-fit values given in Tab.~\ref{mixings}. 
These plots confirm the conclusions drawn from  Eq.~(\ref{pion}).
Although not explicitly shown, the impact of $\theta_{12}$ is completely negligible for $R_{\mu\tau}$, whereas for $R_{e\mu}$, it 
is numerically comparable to that of $\theta_{13}$.

\begin{figure}[h!]
\includegraphics[width=0.45\textwidth]{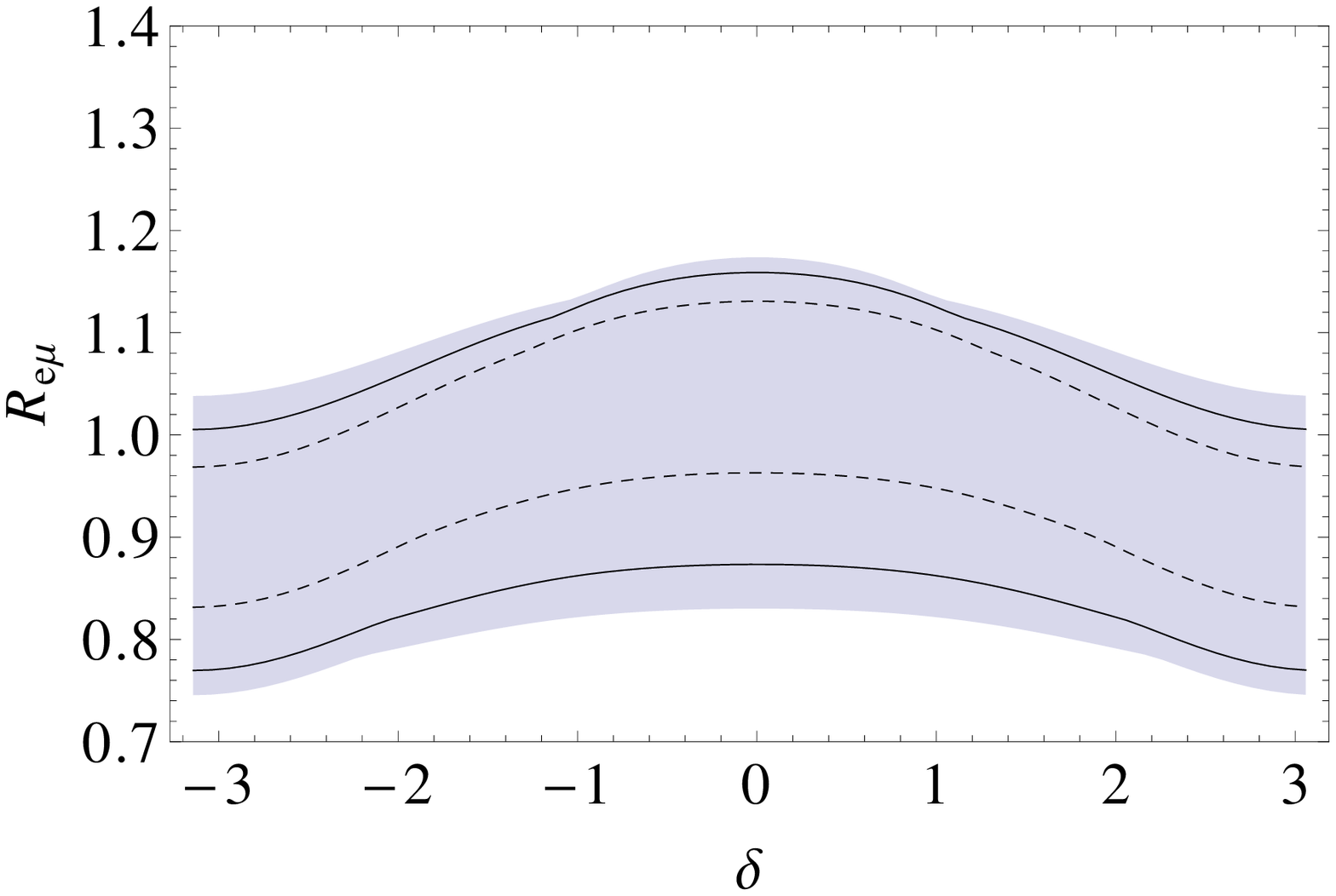}\qquad \includegraphics[width=0.45\textwidth]{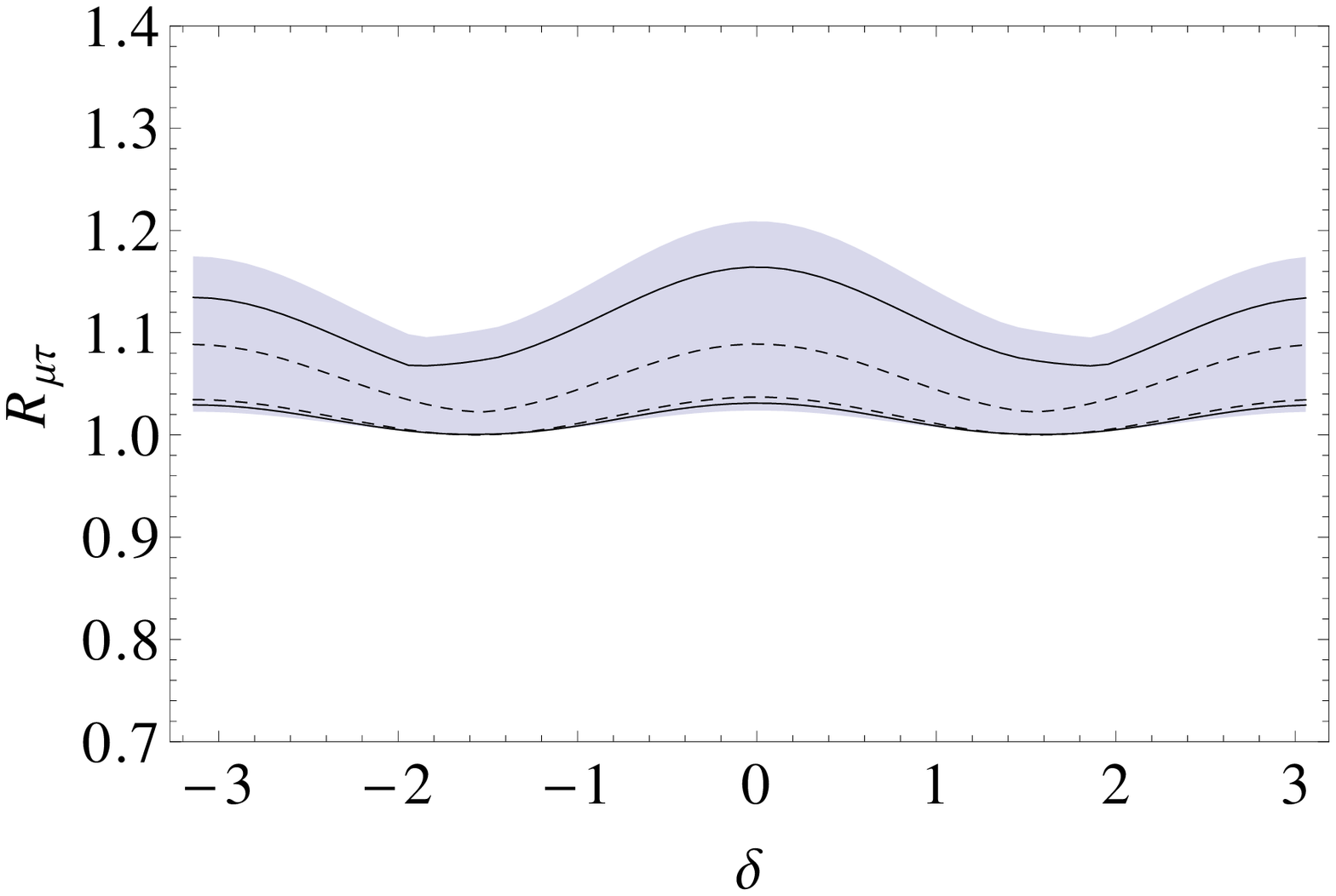} \\
\centering
\includegraphics[width=0.45\textwidth]{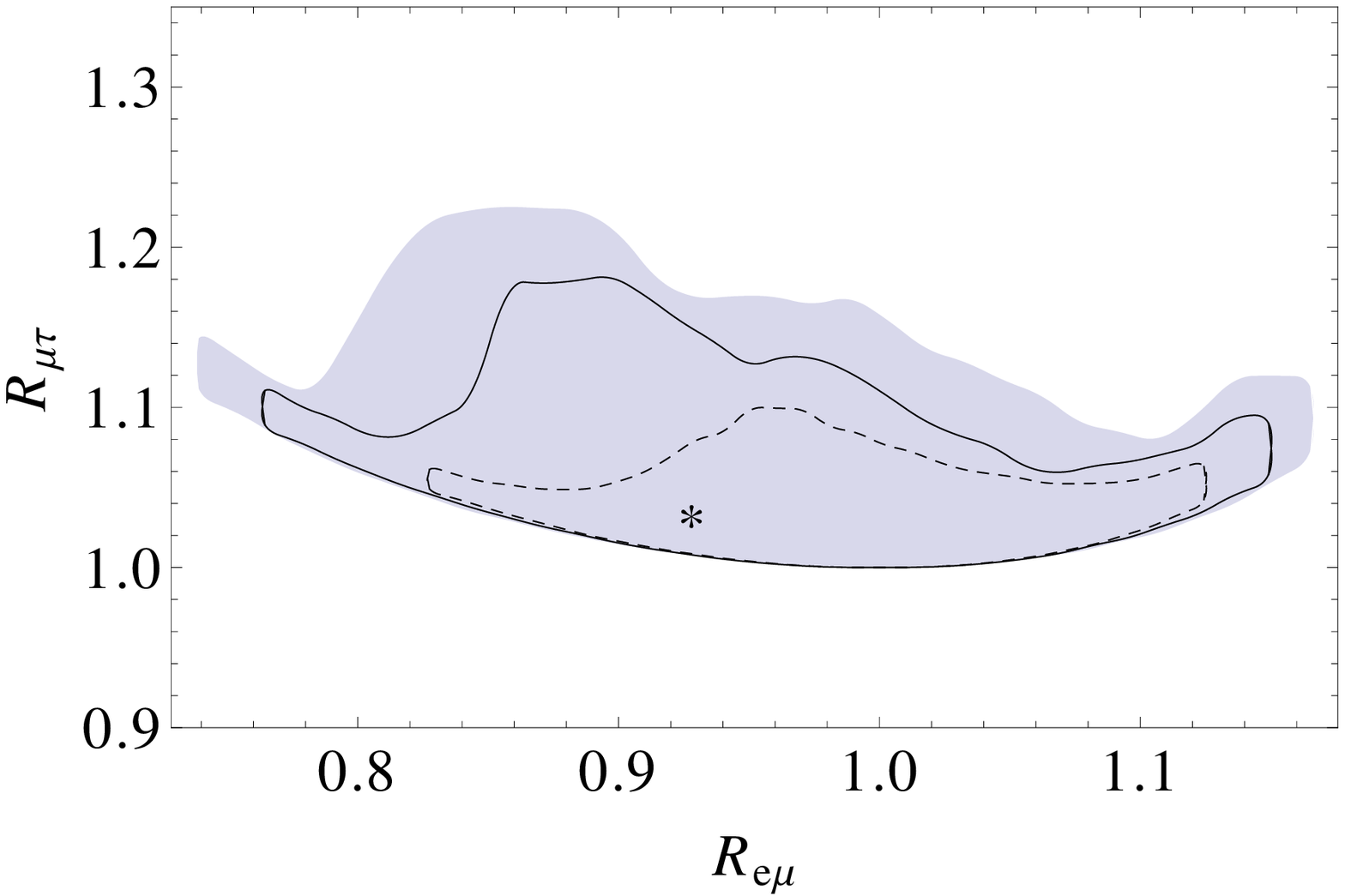} 
\caption{\it Total uncertainties on $R_{e\mu}$ (upper-left plot) and $R_{\mu\tau}$ (upper-right plot) as functions of $\delta$ when all 
other mixing parameters are varied within their 3$\sigma$ errors. In the lower plot, we show the correlation
between $R_{e\mu}$ and $R_{\mu\tau}$. Note that $\delta$ is left free in the interval $[-\pi,\pi]$.}
\label{fig:unc2}
\end{figure}

As a next step, we evaluate all allowed values of the flux ratios considering the simultaneous variations of all parameters within their respective uncertainties as specified in 
Tab.~\ref{mixings}. The results are shown in Fig.~\ref{fig:unc2} (upper plots), where we also show the present correlation between the two flux ratios (lower plot). In these plots, the bands indicate allowed regions where all parameters are varied within their respective 3$\sigma$ ranges, 
whereas the solid and dashed curves have been obtained from the 2$\sigma$ and 1$\sigma$ ranges, respectively.
Figure~\ref{fig:unc2} deserves some more comments. Let us consider the allowed regions within the dashed curves, which can illustrate a possible situation when all mixing parameters (especially $\theta_{23}$) 
will be known with better accuracy than today. In this case, the dependence of the flux ratios on $\delta$ is pronounced, 
since $R_{e\mu}$ has maximum and minimum values 
corresponding to the CP-conserving values $\delta = 0, \pm \pi$, respectively;
at $\delta = \pm \pi$, $R_{e\mu}\in [0.83,0.97]$ and at $\delta = 0$, $R_{e\mu}\in [0.96,1.13]$.
For $R_{\mu\tau}$,  the $\delta$ dependence is more complicated, so that 
the flux ratio assumes similar values in $[1.03,1.09]$ for $\delta = 0,\pm \pi$,
but $R_{\mu\tau}\in [1.00,1.03]$ for $\delta = \pm \pi/2$.

If the errors on the mixing angles were larger, the dependence on $\delta$ would be less pronounced, but still enough to extract information on it (see below).  Globally, we obtained the following ranges for the flux ratios:
\bea
\nn
R_{e\mu}\in [0.85,1.18]\,,\qquad R_{\mu\tau}\in [1.00,1.21]\,.
\eea
The correlation between the two flux ratios shown in the lower plot of Fig.~\ref{fig:unc2} allows one to make consistency checks. Working in the absence 
of exotic processes (like neutrino decays), two  simultaneous measurements of $R_{e\mu}$ and $R_{\mu\tau}$ should fall in these regions. 
For example,  $R_{e\mu}\sim 0.8$ must be accompanied by $R_{\mu\tau}\sim 1.1$ to give  $|\delta| \sim \pi$.
It is interesting to note that the authors of Ref.~\cite{Fogli:2012ua} found a possible hint
in favor of $\delta \sim \pi$. Thus, given all other best-fit values, these results -- if they are confirmed -- would imply 
$(R_{e\mu},R_{\mu\tau})=(0.93,1.04)$, which is indicated by a ``star'' in the lower plot of Fig.~\ref{fig:unc2}.

\subsection{Discussion on muon-damped sources}

\begin{figure}[h!]
\centering
\includegraphics[width=0.6\textwidth]{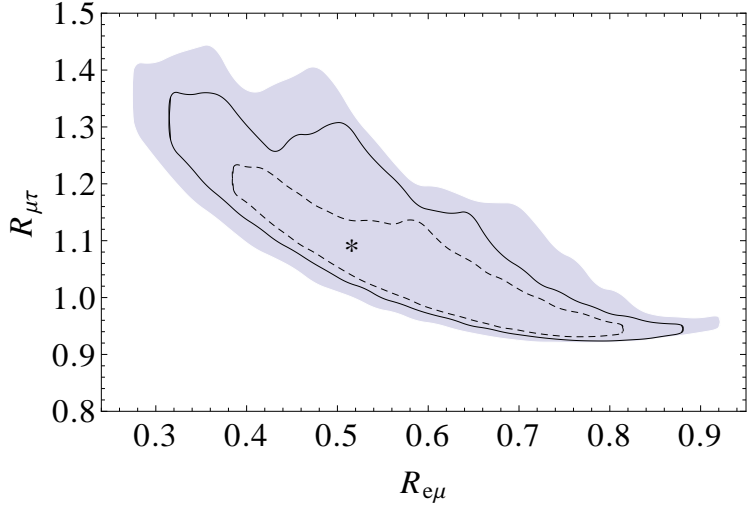}
\caption{\it Correlation between $R_{e\mu}$ and $R_{\mu\tau}$ for the $0:1:0$ case obtained when the mixing parameters vary within their 1$\sigma$, 2$\sigma$, and 3$\sigma$ ranges. Note that $\delta$ is left free in the interval $[-\pi,\pi]$.}
\label{fig:damped}
\end{figure}

A similar study can be carried out for a muon-damped source, for which $\phi^0_{\nu_e}:\phi^0_{\nu_\mu}:\phi^0_{\nu_\tau}=0:1:0$.
Up to first order in the small parameters, the relevant flux ratios read
 \bea
 R_{e\mu} &=& \frac{4\,\sin^2 (2\theta_{12})}{7+\cos (4\theta_{12})} + \frac{64\,\sin (4\theta_{12})}{[7+\cos (4\theta_{12})]^2}\,\delta_{12} - \frac{4\,\cos(\delta)[-9+\cos (4\theta_{12})]\sin (4\theta_{12})}{[7+\cos (4\theta_{12})]^2}\,\theta_{13} \nn \\
 &&
 +
 \frac{8\,[-9+\cos (4\theta_{12})]\sin^2 (2\theta_{12})}{[7+\cos (4\theta_{12})]^2}\,\delta_{23} \nn
 \,,\\  \label{eq:RemRmt} \\ \nn
 R_{\mu\tau} &=& 1 -  \frac{4\,\cos(\delta) \sin (4\theta_{12})}{7+\cos (4\theta_{12})}\,\theta_{13} + \frac{8\,\sin^2 (2\theta_{12})}{7+\cos (4\theta_{12})} \,\delta_{23}\,. \nn
 \eea
Inserting the best-fit values of the mixing parameters, we obtain
\bea
R_{e\mu} &=&  0.6 + 1.1\, \delta_{12} +
 0.7  \,\cos(\delta)\, \theta_{13}- 1.7\, \delta_{23}  + {\cal O}(\delta_{ij}^2)  \nn
\,,\\  \label{eq:RemRmtn} \\ \nn
R_{\mu\tau} &=& 
1.0- 0.4\, \cos(\delta)\, \theta_{13}+ 0.9\, \delta_{23}  + {\cal O}(\delta_{ij}^2)
\,, \nn
\eea
where the uncertainty from $\delta_{12}$ is negligible at this order in $R_{\mu\tau}$.
Some important differences arise when comparing the formulas in Eq.~(\ref{eq:RemRmtn}) with the corresponding ones in Eq.~(\ref{pion}). In fact, $R_{e\mu}$ now obtains a leading dependence 
on $\delta_{12}$, meaning that this flux ratio is more sensitive to the uncertainty of $\theta_{12}$ than the corresponding ratio for the $1:2:0$ case. At the same time, $R_{\mu\tau}$ is corrected by the standard unit value
by linear terms in $\theta_{13}$ and $\delta_{23}$, which means that the fluctuations due to the current uncertainties on the mixing angles 
are more pronounced than before. These considerations are based on analytical formulas that have been verified numerically. As for the previous case, we found that the uncertainty on $\theta_{23}$ is 
the dominant source of error, and for $R_{e\mu}$, the error induced by $\theta_{12}$ is larger than the one given by $\theta_{13}$.
Besides these differences, the qualitative behavior of the flux ratios as functions of $\delta$ is similar to the ones shown in the upper plots of 
Fig.~\ref{fig:unc2}.
On the other hand, the correlation between the two flux ratios is different from the previous case, so we display it in Fig.~\ref{fig:damped}.
As can be observed, there exists a ``negative'' correlation, extending for $R_{e\mu} \in [0.27,0.92]$ and $R_{\mu\tau} \in [0.92,1.42]$.

\subsection{Potential to measure a non-vanishing CP-violating phase}
\label{exp}

Following Ref.~\cite{delta2}, we study the potential of neutrino telescopes to distinguish between CP-conserving and CP-violating values of $\delta$, using the possible measurements of $R$ and $R_{e\tau}$ and also the capability to differentiate between  $\pi S$ and $\mu DS$ sources. This is a difficult task, as was outlined in Ref.~\cite{Lipari:2007su}, especially if 
the uncertainties on the considered flux ratios are large.
For the sake of completeness, we present the dependence on $\delta$ for $R$ and $R_{e\tau}$ computed at the best-fit values of $\theta_{ij}$ for both $\pi S$ and $\mu DS$ sources:
\bea
R^{\pi S}&\sim&\frac{3}{1.99+ 0.03 \cos(\delta)}-1\,,\qquad 
R_{e\tau}^{\pi S}\sim\frac{19.1+\cos(\delta)}{18.7- 0.47 \cos(\delta) - 0.21 \cos(2\delta)}\,, \nn  \\ \label{others} \\
R^{\mu DS}&\sim&\frac{1}{0.61+ 0.03 \cos(\delta)}-1\,,\qquad 
R_{e\tau}^{\mu DS}=\frac{8.94+\cos(\delta)}{14.3+ 0.02 \cos(\delta) - 0.21 \cos(2\delta)}\,.  \nn
\eea
The coefficients of the $\cos(\delta)$ terms are small and relevant for the 
CP-conserving values. In addition, there are no preferred sources in the determination of $\delta$, since the $\delta$ dependence in the flux ratios built from $\pi S$ and $\mu DS$ is not really dissimilar.

\begin{figure}[h!]
\centering
\includegraphics[width=0.55\textwidth]{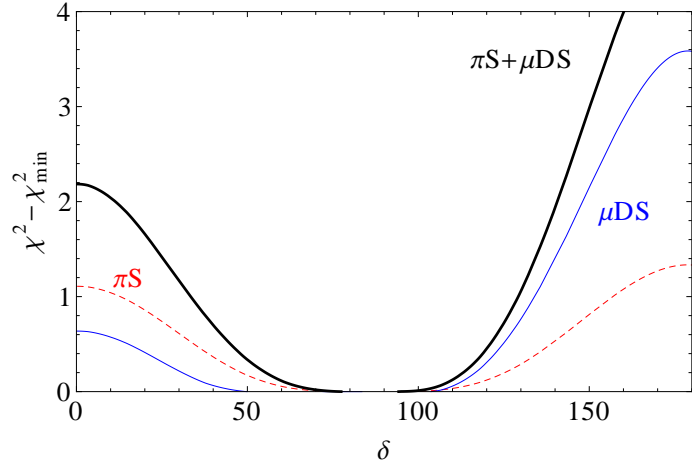}
\caption{\it $\chi^2-\chi^2_{min}$ as a function of $\delta$, after minimizing over all non-displayed parameters. The three curves refer to 
$\pi S$ (dashed curve), $\mu DS$ (thin solid curve), and the sum of both (thick solid curve).}
\label{fig:fit}
\end{figure}

\begin{figure}[h!]
\centering
\includegraphics[width=0.6\textwidth]{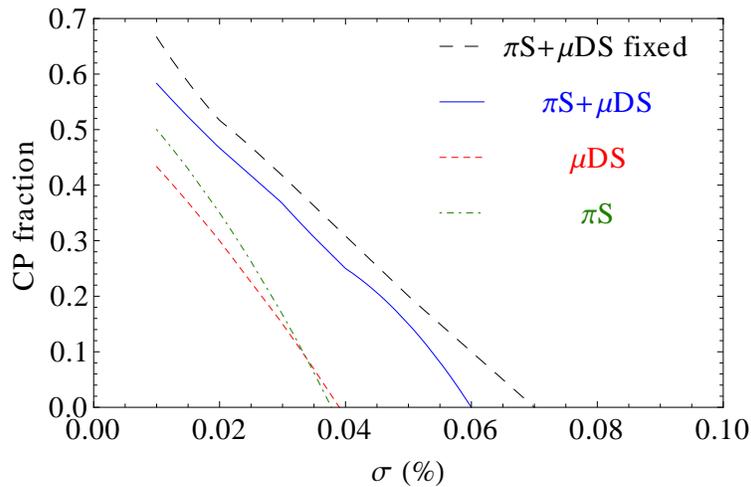}
\caption{\it 2$\sigma$ CP fraction estimate as a function of the common flux ratio uncertainty. 
The four cases correspond to $\pi S + \mu DS$ (solid and long-dashed curves), $ \mu DS$ (dashed curve), and $\pi S $ (dot-dashed curve). }
\label{fig:fit2}
\end{figure}

To evaluate how $\delta=0$ can be distinguished from other non-vanishing values, we construct a $\chi^2$ function
\bea
\label{chi2fun}
\chi^2 = \left[\frac{R^{\rm exp}-R(\theta_{ij},\delta)}{\sigma_R}\right]^2 + 
\left[\frac{R_{e\tau}^{\rm exp}-R_{e\tau}(\theta_{ij},\delta)}{\sigma_{R_{e\tau}}}\right]^2\,,
\eea
where $R^{\rm exp}$ and $R_{e\tau}^{\rm exp}$ are $R$ and $R_{e\tau}$ evaluated at the 
best-fit points for $\theta_{ij}$ as given in Tab.~\ref{mixings}. We consider the possibility of having the sum of both contributions from $\pi S$ and $\mu DS$ sources. In the minimization procedure, we marginalize over all parameters but $\delta$.
As an illustrative example, we show in Fig.~\ref{fig:fit} the case where $\delta=\pi/2$.
In the $\chi^2$ function, we assume 5~\% error on the flux ratios.
As expected, the combination of both $\pi S$ and $\mu DS$ sources gives the best resolution for $\delta$.
In particular, the $\chi^2$ function does not touch the CP-conserving values $\delta=(0,\pi)$ at 2$\sigma$, 
so maximal CP violation can be distinguished from CP conservation. 
This can be repeated for every input value of $\delta$; at a given confidence level and 
flux ratio uncertainty, there exist a range of phases for which the $\chi^2$ function does not touch $\delta=(0,\pi)$;
the fraction of such points over the whole $[0,\pi]$ range defines a {\it CP fraction} that serves to illustrate the goodness of the neutrino telescope to access CP violation. The result is reported in 
Fig.~\ref{fig:fit2}, where we display the 2$\sigma$ CP fraction in terms of the uncertainty $\sigma_R = \sigma_{R_{e\tau}}=\sigma$, 
ranging from 1~\% to 10~\%, obtained using the $\chi^2$ function defined in Eq.~(\ref{chi2fun}). 
We have considered four cases; the long-dashed curve refers to the case in which all mixing parameters but $\delta$ are 
fixed to their best-fit values and all neutrino sources are taken into account.
It is clear that the best performance is reached when the maximum amount of information is collected (i.e., the sum of the sources). However, such performances are limited to $\sigma \lesssim 0.065$. Even considering infinite precision on the mixing angles, the CP fraction is  non-vanishing only for $\sigma \lesssim 0.07$. This is the consequence of the mild $\delta$-dependence on the flux ratios in Eq.~(\ref{others}). Nevertheless, since astrophysical high-energy sources have not been observed so far, a more promising possibility would be to combine future neutrino telescope data with other future experimental data (from e.g.~NOvA), which should have some sensitivity to the octant of $\theta_{23}$, and thus reducing the intrinsic uncertainties on the flux ratios.

\section{Summary and conclusions}
\label{sec:sc}

In this paper, we have analyzed the potential of neutrino telescopes to access the leptonic CP-violating phase $\delta$. We have derived expansions for the flux ratios $R_{\alpha\beta}$ 
up to first (and second) order in small parameters, explicitly showing their dependence on $\delta$, for both kinds of high-energy neutrino sources, i.e., $\pi S$ and $\mu DS$ sources. It turns out that 
the uncertainty on $\theta_{23}$ affects the global (theoretical) error on 
the flux ratios the most.  Considering both kinds of sources, we have shown 
that a 10~\% of CP fraction can still be obtained with a 5.5~\% uncertainty on $R_{\alpha\beta}$.
We urge IceCube to measure the flux ratios, since such a measurement could provide the first hints on the value of $\delta$.

\section*{Acknowledgments}

This work was supported by the program ``Futuro in Ricerca 2010 (RBFR10O36O)'' (D.M.), the Swedish Research Council (Vetenskapsr{\aa}det), contract no. 621-2011-3985 (T.O.), and the G{\"o}ran Gustafsson Foundation (D.M.).
D.M. would also like to thank for the hospitality of the KTH Royal Institute of Technology, where part of this work was performed.

\end{document}